%% file: main.tex
\documentclass[%
 reprint,
 superscriptaddress,
 amsmath,amssymb,
 aps,
]{revtex4-2}

\usepackage{graphicx}
\usepackage{dcolumn}
\usepackage{bm}
\usepackage{mathrsfs}
\usepackage{comment}
\usepackage{float}
\usepackage{xcolor} 
\usepackage{booktabs}

\begin{document}

\title{Ethics in rotten apples:\\ A network epidemiology approach for active cyber defense}

\author{Francesco Bonacina} \affiliation{INSERM, Sorbonne Université, Pierre Louis Institute of Epidemiology and Public Health, Paris, France} \affiliation{Sorbonne Université, CNRS, Laboratoire de Probabilités, Statistique et Modélisation, LPSM, Paris, France}

\author{Ignacio Echegoyen}

\affiliation{Grupo Interdisciplinar de Sistemas Complejos (GISC) \& Departmento de Psicología, Universidad Pontificia Comillas, 28049 Madrid, Spain}

\author{Diego Escribano}

\affiliation{%
 Grupo Interdisciplinar de Sistemas Complejos (GISC), Departamento de Matem\'aticas, Universidad Carlos III de Madrid, Legan\'es, Madrid, Spain
}

\author{Marcus Krellner}

\affiliation{Teesside University, UK}

\author{Francesco Paolo Nerini}

\affiliation{
 CENTAI Institute, Turin, Italy
}

\author{Rasha Shanaz}
\affiliation{%
 Department of Physics, Bharathidasan University, Tiruchirappalli, India
}%

\author{Andreia Sofia Teixeira}
\affiliation{%
LASIGE, Departamento de Informática, Faculdade de Ciências, Universidade de Lisboa, Lisboa, Portugal
}%

\author{Alberto Aleta}
\affiliation{%
 Institute for Biocomputation and Physics of Complex Systems, University of Zaragoza, Zaragoza, Spain
}%

\date{\today}

\begin{abstract}

\input{files/0_abstract.tex}

\end{abstract}

\maketitle


\section{\label{sec:intro}Introduction}
\input{files/1_introduction.tex}

\section{\label{sec:materials_methods}Materials and methods}
\input{files/2_materials_methods.tex}

\section{\label{sec:results}Results}
\input{files/3_results.tex}
\section{\label{sec:discussion_conclusions}Discussion and conclusions}
\input{files/4_discussion_conclusions.tex}

\input{files/5_acknowledgment.tex}

\appendix
\input{files/6_appendix_gillespie.tex}

\input{files/7_appendix_mean_field.tex}

\bibliography{references_new}

\end{document}

%% file: files/0_abstract.tex
As Internet of Things (IoT) technology grows, so does the threat of malware infections. A proposed countermeasure, the use of benevolent ``white worms'' to combat malicious ``black worms'', presents unique ethical and practical challenges. This study examines these issues via network epidemiology models and simulations, considering the propagation dynamics of both types of worms in various network topologies. Our findings highlight the critical role of the rate at which white worms activate themselves, relative to the user's system update rate, as well as the impact of the network structure on worm propagation. The results point to the potential of white worms as an effective countermeasure, while underscoring the ethical and practical complexities inherent in their deployment.

%% file: files/1_introduction.tex
`Internet of Things' (IoT) technology is everywhere. Even seemingly trivial household devices like light bulbs and toasters are connected to the internet over local networks. Unfortunately, the rise of malware infections has become a critical concern in IoT cybersecurity, posing a significant threat with increasing frequency and sophistication. These infections lead to disruptive system failures and substantial financial losses \cite{Sengupta2020Jan, sinanovic}. In response, a promising countermeasure has emerged in the form of “white worms”, which would serve as benevolent counterparts to malicious “black worms”.

In this context, worms refer to a type of malware that exploits vulnerabilities in devices to propagate to other devices. Unlike smartphones and personal computers, IoT devices typically lack regular updates \cite{molesky_internet_2019}. The proposed white worms share similar propagation characteristics with black worms but are specifically designed to identify and rectify security vulnerabilities \cite{de_donno_antibiotic_2018, de_donno_combining_2019}.

However, before white worms can be widely adopted, there are significant questions that must be addressed. Ethically, the concept of white worms walks a fine line since they infiltrate systems without explicit permission, which could be viewed as a breach of privacy or even illegal. This raises intricate ethical and legal dilemmas that require careful exploration, potentially limiting the application of white worms. Additionally, understanding the propagation dynamics of these worms is vital for designing effective and ethical white worms.

The propagation of viruses, whether biological or digital, has been a focal point of scientific investigation for many decades. As early as the 1980s, there have been propositions that computer viruses could be studied using tools and methodologies developed for human diseases \cite{Murray1988Apr}. The tools developed by network epidemiology are particularly suited to this task given the resemblance to biological networked systems, and the mechanisms by which viruses spread \cite{Kephart1991May, Pastor-Satorras2015Aug, Pastor-Satorras2001Apr,berger_spread_nodate,yom-tov_modeling_2019}.

While the spread of multiple viruses on networks has been studied in network epidemiology \cite{bernini_evaluating_2019, millerCocirculationInfectiousDiseases2013}, we propose a model specifically tailored to the contagion of computer viruses, wherein one of the pathogens protects the host from further infection. Furthermore, we incorporate the ethical characteristics of white worms proposed in the literature \cite{de_donno_antibiotic_2018, de_donno_combining_2019}. To accomplish this, we develop a compartmental model that spreads on various types of networks and explore its dynamics through stochastic simulations under different conditions. Finally, we discuss the effectiveness of white worms, considering the ethical considerations incorporated into the model.

%% file: files/2_materials_methods.tex
\subsection{Overview of the contagion process}

Our model considers the propagation of two worm types within a network of vulnerable devices ($V$): a malicious “black worm” and a benign “white worm”. The black worm's purpose is to infiltrate any unprotected device by exploiting an unspecified security loophole, with a transmission rate $\beta_B$ from one device to another. Conversely, the white worm seeks to secure the devices by forcing system updates. We label its transmission rate $\beta_W$ and hypothesize that both types of worms exploit the same security loophole, equating their transmission rates, i.e., $\beta_W = \beta_B$. However, in line with the suggestion made by \cite{de_donno_antibiotic_2018}, the white worm does not take immediate action upon the device. Initially, it urges the device's user to update the system while remaining in a dormant state ($D$). The user has the option to patch the system's vulnerability at a rate of $\gamma$.

Subsequently, the white worm uses the device's resources to (i) propagate to connected machines and (ii) patch the system. Between the period of activation and updating (states $W$ or $W_B$), the white worm maintains the capacity to spread, but the user has no possibility to update the device manually. It is important to note that the mere presence of a worm does not eliminate the device's vulnerability. Therefore, a device hosting a dormant or active white worm can still be compromised by the black worm ($D_B$ or $W_B$). Similarly, a device already infected by the black worm ($B$) can be infiltrated by the white worm ($D_B$). The white worm transitions from a dormant to an active state at a rate $\epsilon$. Once in the active state, it initiates the system update at a rate $\mu$, hence sealing the security loophole. Once protected, white and black worms are removed, and the device becomes immune to further infections by any of them ($P$).

\begin{figure}[H]
    \centering
    \includegraphics[width=\linewidth]{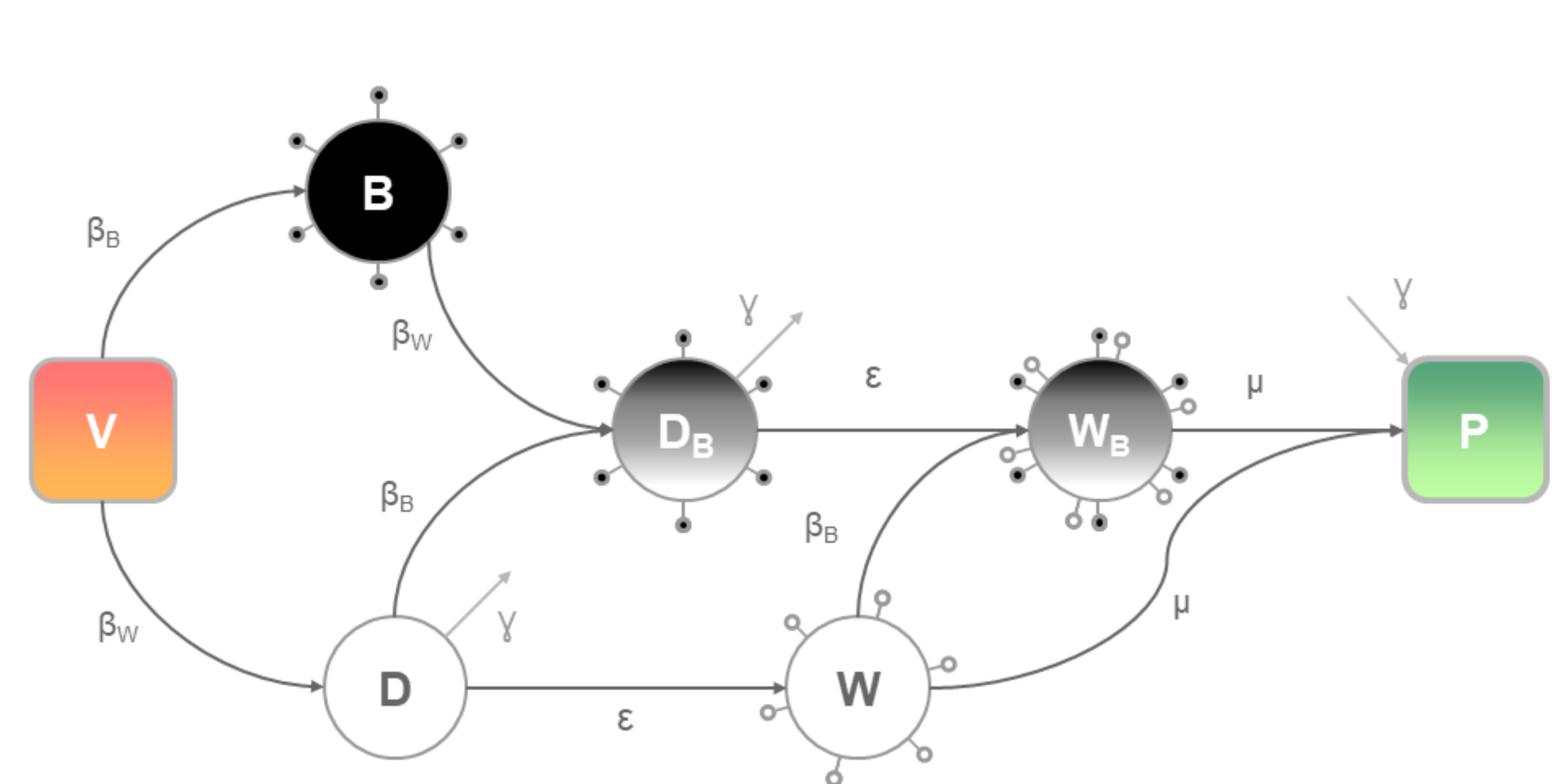}
    \caption{\textbf{Model scheme} - Compartmental model that describes how black and white worms can spread within the system. Vulnerable devices ($V$) can be infected by either a black worm or a white worm. When infected by a black worm ($B$), they actively spread it. However, if infected by a white worm, they enter a dormant state ($D$) until system upgrade or self-activation occurs. Activated white worms ($W$) propagate until the device is forcibly updated. Devices with white worms in dormant or active states can also be infected by black worms ($D_B$ or $W_B$). Similarly, devices infected with black worms ($B$) can be infected by active white worms ($D_B$). Once the system is updated, either by user approval or the action of a white worm, security vulnerabilities are fixed, and the machine is protected ($P$).
    }
    \label{fig:VIPmodel}
\end{figure}

The diagram depicted in Fig.~\ref{fig:VIPmodel} represents all possible state transitions within the system.

\begin{table}[h]
\centering

\begin{tabular}{lcc}
\toprule
\textbf{Name} & \textbf{Parameter} & \textbf{Value} \\
\midrule
Infection rate of black worms & $\beta_B$ & 1.1 \\
Infection rate of white worms & $\beta_W$ & 1.1 \\
Activation rate of white worms & $\epsilon$ & [0.01-1000]\\
Protection rate (user) & $\gamma$ & [0.1-100]\\
Protection rate (white worm)  & $\mu$ & 1 \\
\bottomrule
\end{tabular}
\caption{\label{tab:transition-parameters}
\textbf{Summary of the transmission parameters of the model} - For both black and white worms, the infection rate parameters are set equal. Furthermore, without loss of generality, the protection rate associated with the white worm is also fixed. Lastly, we will iterate the activation rate values for white worms and the user's protection rate within the intervals specified in the table. }

\end{table}

In accordance with common practice, we set $\mu = 1$ without losing generality, as time can always be appropriately rescaled. We are primarily interested in the scenario where both worm types exploit the same vulnerability for propagation, hence $\beta_B = \beta_W$, as previously established. Consequently, our analysis concentrates on the influence of two parameters that pertain to the ethical conduct of the white worm: the rate $\epsilon$ at which a dormant white worm is activated and starts to leverage the resources of the host device, and the rate $\gamma$ at which users respond to system update prompts.
A summary of the transmission parameters explored in this study is described in Table \ref{tab:transition-parameters}.

\subsection{Epidemic dynamics in the homogeneous mixing}

Let us define $\rho^X(t)$ as the fraction of devices in the compartment $X$ at time $t$, i.e., its density. Then, the equations of the model under the homogeneous mixing assumption are the following:
\begin{equation}\label{eq:homo}
\begin{cases}
&\dot{\rho}^V = - \beta_B \rho^V \phi^B - \beta_W \rho^V \phi^W,  \\
&\dot{\rho}^B =  \beta_B \rho^V \phi^B - \beta_W \rho^B \phi^W, \\
&\dot{\rho}^D = \beta_W \rho^V \phi^W - \beta_B \rho^D \phi^B- \epsilon \rho^D - \gamma \rho^D, \\
&\dot{\rho}^{D_B} = \beta_B \rho^D \phi^B + \beta_W \rho^B \phi^W - \epsilon \rho^{D_B}  - \gamma \rho^{D_B},  \\
&\dot{\rho}^W = \epsilon \rho^D - \beta_B \rho^W \phi^B - \mu \rho^W, \\
&\dot{\rho}^{W_B} = \epsilon \rho^{D_B} + \beta_B \rho^W \phi^B - \mu \rho^{W_B}, \\
&\dot{\rho}^P = \mu \rho^{W_B} + \mu \rho^W + \gamma \rho^{D_B} + \gamma \rho^D, 
\end{cases}
\end{equation}
where 
\begin{equation}
\begin{cases}
\phi^B &= \rho^B +\rho^{D_B} + \rho^{W_B}, \\
\phi^W &= \rho^W + \rho^{W_B},
\end{cases}
\end{equation}

\noindent
represent the total fraction of devices that can propagate the black or the white worm, respectively.

\subsection{Networks}

Our study scrutinizes worm propagation across three distinct network topologies. The first of these is a complete graph of a hundred nodes. This selection allows us to compare numeric solutions derived from the homogeneous mixing model with results from stochastic simulations. 

However, the structure of real-world computer networks is often far from homogeneous, especially in the case of IoT devices. These networks are known to demonstrate substantial heterogeneity and a high degree of clustering around central access points \cite{Sohn2017Jan,Wan2020Sep}. To better represent this reality, we also consider two different projected network topologies. These projections assume that if two IoT devices are linked to routers that can communicate with each other, a direct link between both devices can be inferred.

To construct these additional network topologies, we employ Python's \texttt{NetworkX} package \cite{osti_960616}. The first of these is an Erd\H{o}s-R'enyi network, where pairs of nodes establish connections with a consistent probability. The second network follows a power law distribution, with node degrees $k$ conforming to a power law distribution, $k^{-\alpha}$, where $\alpha = 10$. The contagion process equations for a network under mean-field approximation are provided for further insight in Appendix \ref{app:meanfield}.

\subsection{Stochastic simulations}

The stochastic propagation of both worm types across the network is simulated utilizing the Gillespie algorithm~\cite{gillespie_general_1976}. Originally proposed as a Monte Carlo simulation technique for chemical reactions, it has since been extended to model Markovian dynamics, such as those observed in epidemics \cite{cai_solving_2016,kiss_mathematics_2017}. More specifically, we have employed the algorithm's implementation found in Python's package \texttt{EoN} version 1.1 \cite{miller_eon_2019,kiss_mathematics_2017}. Detailed insights into this method can be found in Appendix \ref{sec:gillespie}.

%% file: files/3_results.tex
\subsection{Homogeneous mixing}

\begin{figure}
    \centering
    \includegraphics[width=\linewidth]{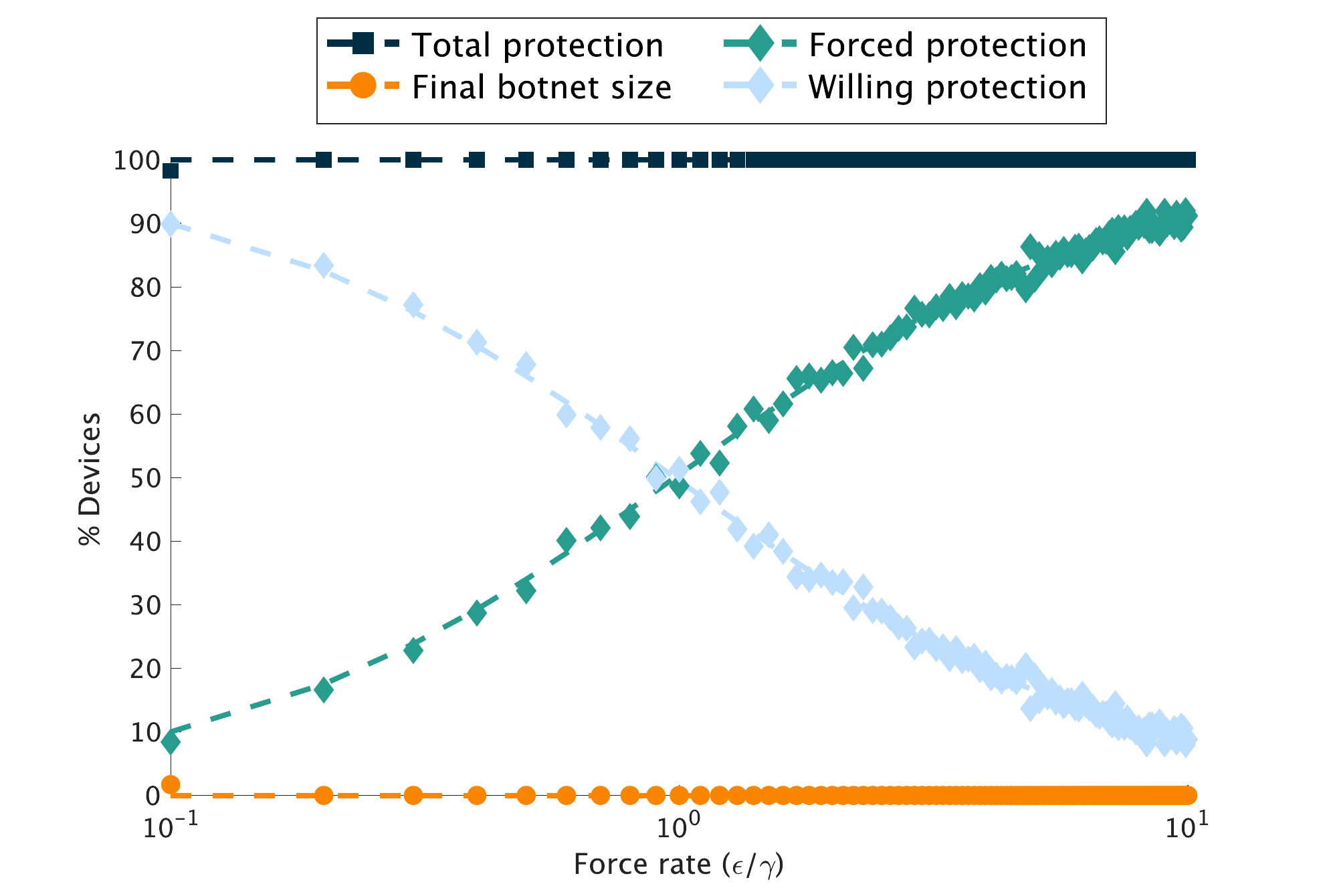}
    \caption{\textbf{Homogeneous mixing} - Final distribution of protected and unprotected devices under the homogeneous mixing hypothesis  as a function of the ratio $\epsilon/\gamma$. We examine the final proportion of protected devices (black), divided into protected by users (green) and protected by the white worm activation (light blue), and the final coverage of the botnet (orange). The results are obtained from both the Equations \eqref{eq:homo} (dashed lines) and the stochastic model implemented on the complete graph (dots).
    }
    \label{fig:Fig2}
\end{figure}

We commence our exploration by examining the model's behavior under the assumption of homogeneous mixing, according to which every device can directly interact with any other device. In Fig.~\ref{fig:Fig2}, we depict the final proportion of protected devices as a function of the ratio $\epsilon/\gamma$. The majority of observables rely solely on the ratio $\epsilon/\gamma$ and not on their individual values, as altering the value of $\gamma$ only affects the temporal dynamics but not the end states. The four observables depicted in Fig.~\ref{fig:Fig2} were calculated using both the Equations \eqref{eq:homo} and the simulations of the stochastic model.

As we can see in the figure, the major impact of increasing the ratio is changing the path through which devices get protected. If the rate at which users update their system upon being prompted is large ($\epsilon/\gamma \ll 1$), most devices will be protected by their owners. If, instead, the white worm is allowed to spread for a long time before protecting the system ($\epsilon/\gamma \gg 1$), it will actively protect most devices. The stochastic simulations on the complete graph corroborate this finding, showing that the botnet can easily be destroyed under the homogeneous mixing hypothesis.

\subsection{Spreading on networks}

\begin{figure*}
    \centering
    \includegraphics[width=0.9\textwidth]{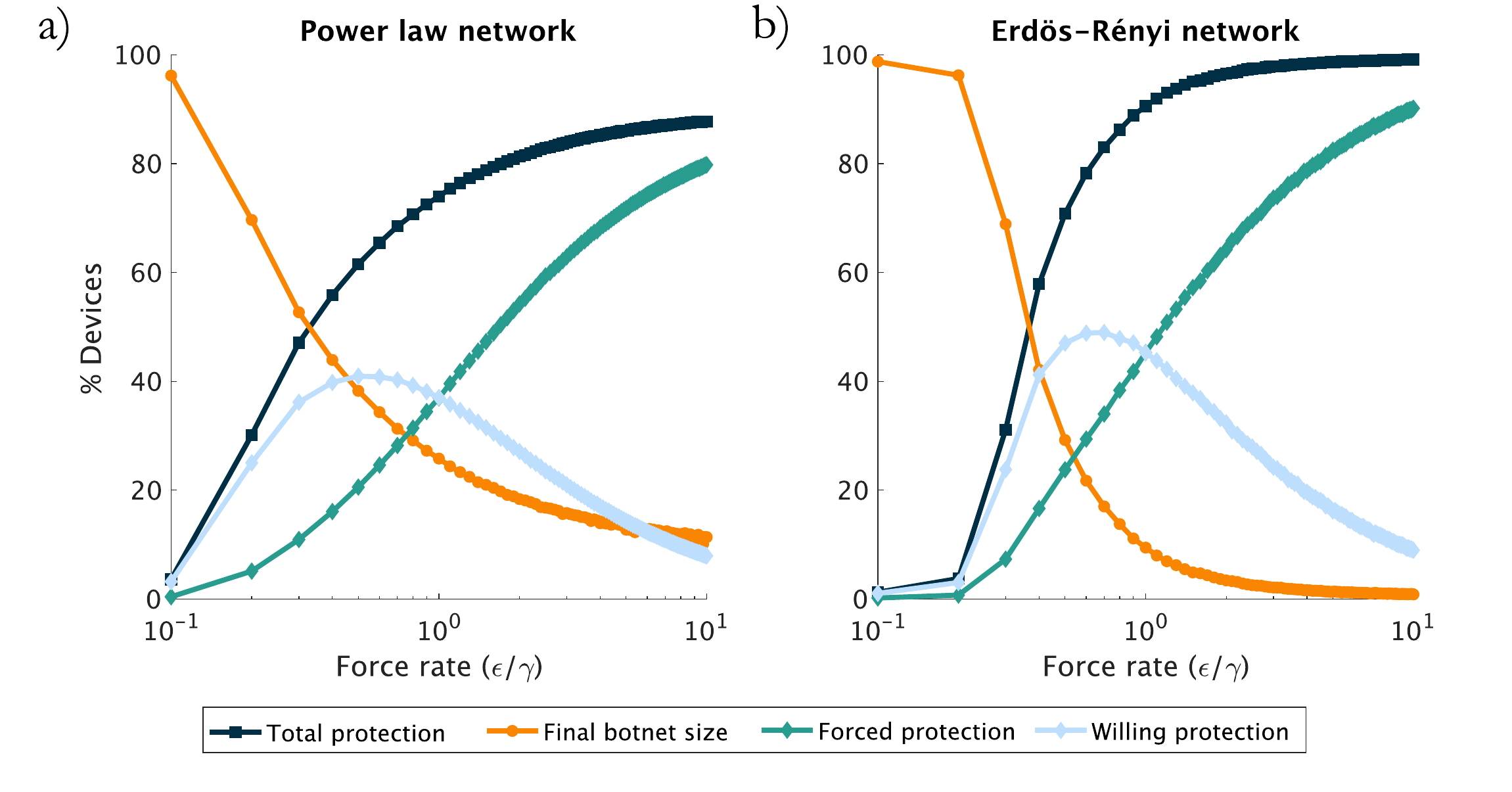}
    \caption{\textbf{Protection coverage for spreading on networks} - Total fraction of devices protected in the system by the time the white worm vanishes in a power law network (a) and an Erd\H{o}s-R\'enyi network (b) as a function of the force rate ($\epsilon/\gamma$). We distinguish whether the protection was provided by a willing update by the owner of the device (light blue) or forced by the white worm (green). In orange, the final botnet size by the end of the simulation. All results were obtained using stochastic simulations of the model with the parameters described in Table~\ref{tab:transition-parameters}.}
    \label{fig:Fig3}
\end{figure*}

Under the homogeneous mixing model, the final size of the botnet is essentially zero across a wide range of $\epsilon/\gamma$ values. However, when the spread occurs over heterogeneous networks, the dynamics shift markedly, as illustrated in Fig.~\ref{fig:Fig3}. 

Firstly, we observe the familiar epidemic threshold widely discussed in the relevant literature, which is negligible for scale-free networks \cite{Pastor-Satorras2001Apr}. Consequently, as shown in Fig.~\ref{fig:Fig3}(a), the black worm cannot entirely infect the network, as a significant outbreak of the white worm always leads to the protection of a certain fraction of devices. Contrarily, Fig.~\ref{fig:Fig3}(b) shows that the ratio $\epsilon/\gamma$ must exceed a certain value for the white worm to propagate and dismantle the botnet effectively.

Secondly, we note that even for very high force rates, the final botnet size may not reach zero. In fact, for the power law topology, the botnet size remains over 20\% of the devices even after the white worm's elimination. This finding sharply contrasts with the results from the Erd\H{o}s-R\'enyi network and the complete graph. Moreover, forced device protection by the white worm is required for most scenarios in which the final size of the botnet is relatively small.

\begin{figure*}
    \centering
    \includegraphics[width=0.875\textwidth]{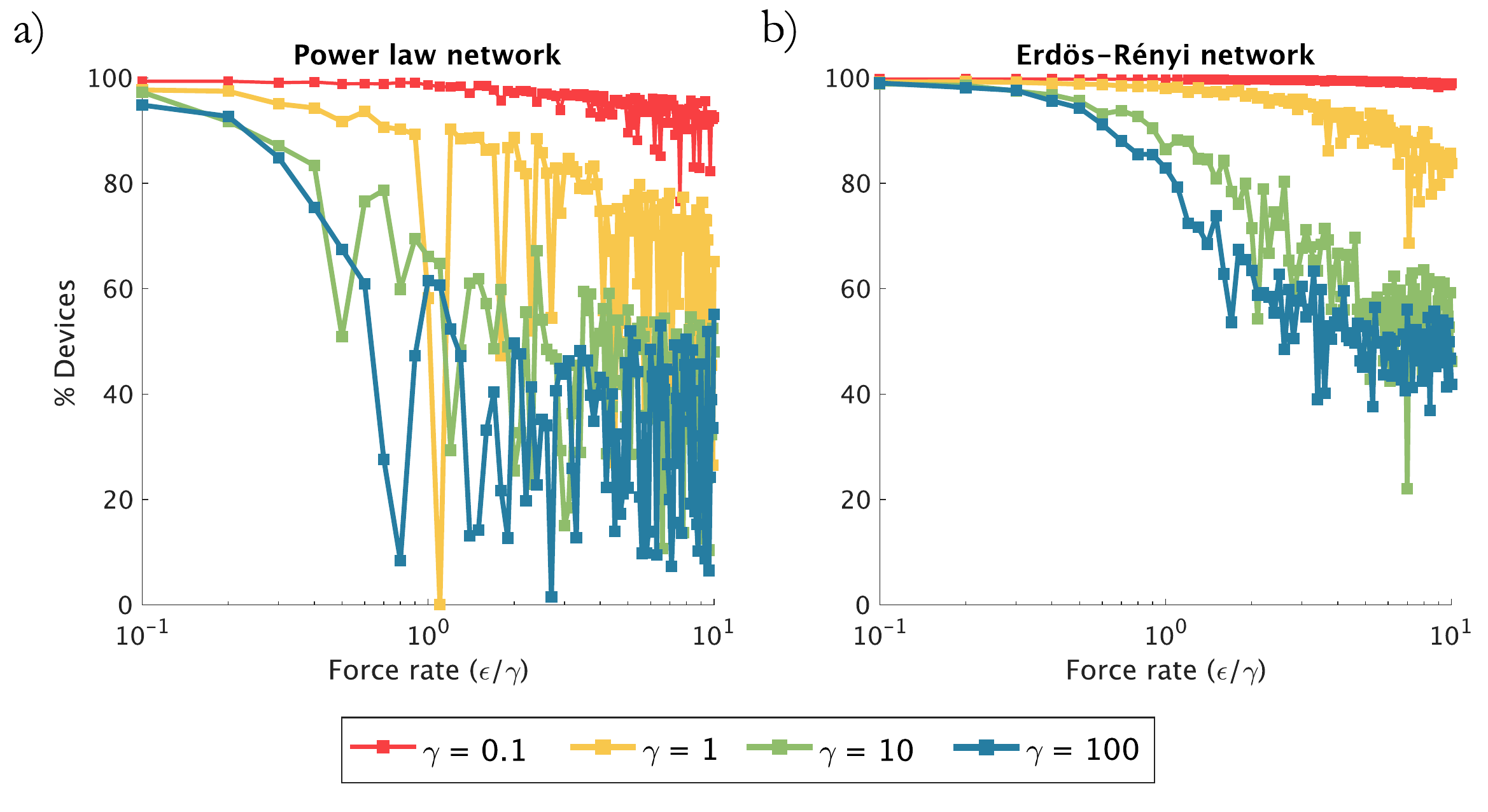}
    \caption{\textbf{\textbf{Maximum botnet size}} - Maximum size reached by the botnet as a function of the force rate $\epsilon/\gamma$ for a power law network (a) and an Erd\H{o}s-R\'enyi network (b). In contrast with other observables, the maximum fraction of devices that simultaneously belong to the network depends on $\gamma$.}
    \label{fig:Fig4}
\end{figure*}

However, these observations merely describe the system's final state and do not consider the dynamics during the initial propagation stages. In Fig.~\ref{fig:Fig4}, we present the fraction of devices that at any point were part of the botnet, i.e. that were simultaneously infected with the black worm and thus exploitable, for instance, for a DDoS attack. Here, the outcomes heavily depend on the specific value of $\gamma$. When its value is exceedingly low, the botnet could potentially cover nearly the entire system at some point. Addressing this issue requires increasing the rate at which users update their systems, as this action is executed much faster than the protection afforded by the white worm. This results in smaller botnets and, consequently, reduced threats, but also requires faster propagation by the white worm (increased $\epsilon$).

\begin{figure}
    \centering
    \includegraphics[width=\linewidth]{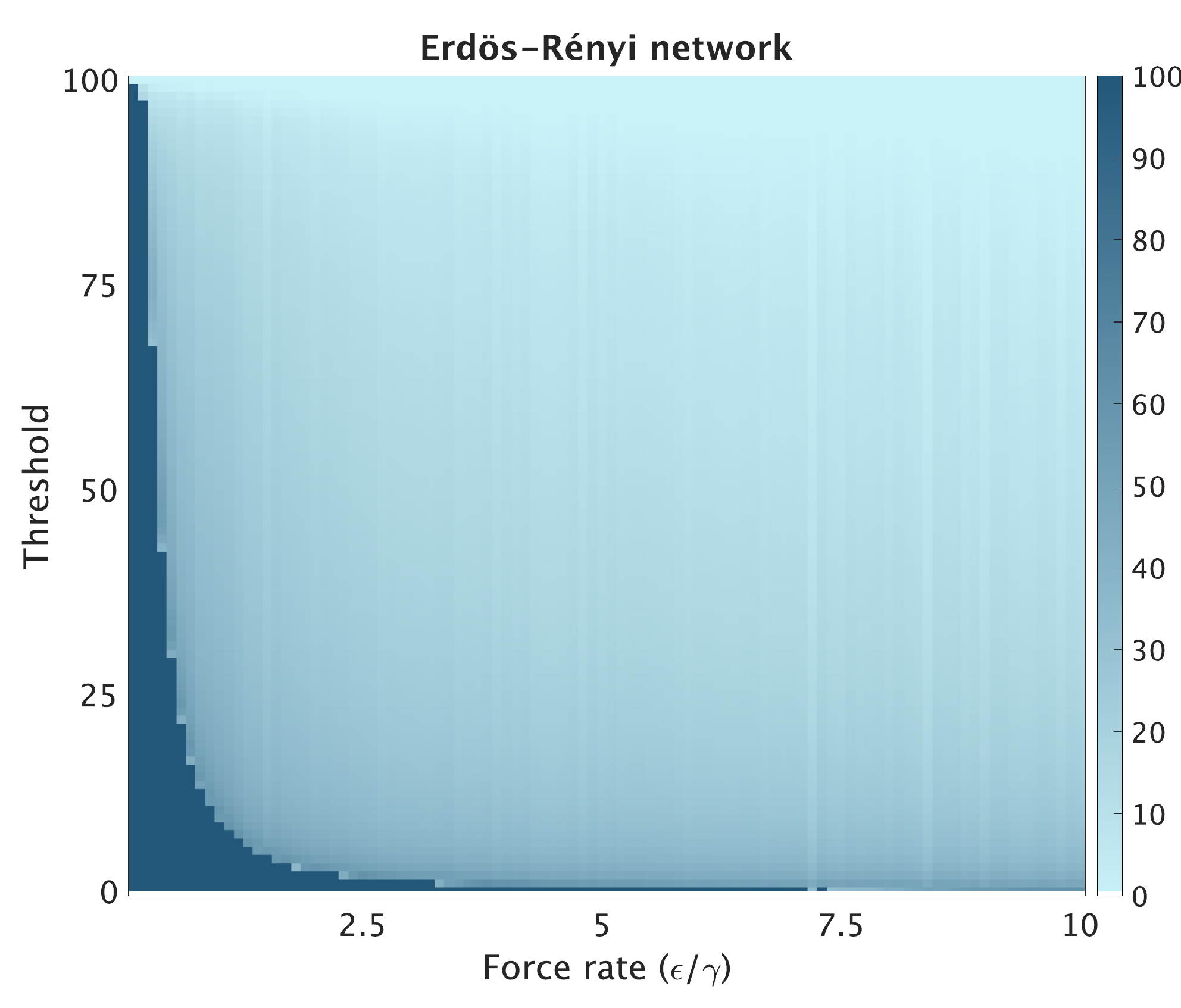}
    \caption{\textbf{Simulation time spent above critical size of the botnet} - Percentage of simulation time spent with the botnet size above different threat thresholds (y-axis), for a range of rates $\epsilon/\gamma$ (x-axis), on a Erd\H{o}s-R\'enyi network. When the system ends with the size of the botnet above a threshold, we set the corresponding active time as $100\%$.} 
    \label{fig:Fig5}
\end{figure}

We conclude this analysis by examining the botnet threat duration. Fig.~\ref{fig:Fig5} portrays the time interval during which the botnet infects a certain fraction of devices as a function of the ratio $\epsilon/\gamma$. We note that when $\gamma \gg \epsilon$, the white worm cannot propagate effectively, and the botnet remains undestroyed indefinitely. However, as we augment $\epsilon$, the interval drastically shortens, thereby reducing the botnet's threat as it can not be used for an extended period of time.

%% file: files/4_discussion_conclusions.tex
The intersection of cybersecurity and ethics presents complex and intriguing dilemmas. Our study attempted to address these challenges through the lens of IoT security and the use of white worms for protection. The introduction of white worms into a system inherently walks a thin ethical line, due to the potential breach of privacy or even legality caused by their self-propagation without explicit user consent. Our findings illuminate both the possibilities and pitfalls that may arise with their use.

In the case of a homogeneous mixing model, we found that for a wide range of the ratio $\epsilon/\gamma$, the botnet is effectively eliminated. Yet, the mechanism leading to its eradication is very different. If $\epsilon \ll \gamma$ the devices are mostly protected actively by their owners. If, instead, $\epsilon \gg \gamma$, the devices will be protected by the white worm. However, the dynamics change significantly when worms spread across heterogeneous networks. We found that in certain cases, the botnet size was never reduced to zero, and over 20\% of devices remained infected in the power law topology. These findings underscore the importance of considering network structure when designing strategies for white worm deployment.

Moreover, our analysis of the early propagation stages revealed that the specific value of $\gamma$ has a significant impact on the size of botnets. A low rate could allow a botnet to cover almost the entire system at some point, underscoring the necessity of taking swift action to protect the system. In other terms, if users do not actively protect their system upon being prompted, the malware will capture most of the system. Furthermore, even if for a wide range of values the complete botnet only lasts for a brief period of time, the fact that malware spreads through the whole system raises other concerns, such as potential data loss or privacy breaches.

Despite these insights, our study has several limitations. Firstly, we made several assumptions about the behavior of white worms and users, which may not hold in real-world situations. For instance, we assumed that both the white and black worms exploit the same security vulnerability. We also assumed that system updates completely protect devices from infection, which may not always be the case given the myriad of potential vulnerabilities. Furthermore, our models do not account for the potential interaction between both worms, such as the white worm directly patching the system if it detects the presence of the black worm.

Future research could address these limitations by incorporating more realistic assumptions and behaviors into the models. Additionally, further empirical studies are necessary to validate the model predictions and to provide more detailed insights into the interactions between white worms, black worms, and users. Similarly, it would be important to study the problem in more realistic IoT networks, as we have seen that the topology plays a major role in the dynamics of the worms.

In conclusion, our study sheds light on the potential of white worms as a countermeasure against black worms in IoT networks. While this strategy could be effective under certain conditions, its implementation raises complex ethical and practical issues that warrant careful consideration. In particular, we have observed that very swift action is necessary, either by the prompted user or directly by the worm, to prevent the creation of a large botnet. This, however, implies that the white worm cannot be ethical (in the sense proposed by \cite{de_donno_antibiotic_2018}) for too long. Further research is needed to fully understand the dynamics of this intriguing interplay between cybersecurity, technology, and ethics.

%% file: files/5_acknowledgment.tex
\section*{Data \& Code}

The code for the Gillespie algorithm, along with the code to generate the networks and solve the model under the homogeneous mixing assumption, is publicly available at \url{https://github.com/FrappaN/C72h-whiteworms}.

\section*{Acknowledgements}

This work is the output of the Complexity72h workshop, held at IFISC in Palma, Spain, 26-30 June 2023, \url{https://www.complexity72h.com}. RS acknowledges the support of SERB International Travel Support grant ref. ITS/2023/001976. AST acknowledges support by FCT -- Fundação para a Ciência e Tecnologia -- through the LASIGE Research Unit, ref.\ UIDB/00408/2020 and ref.\ UIDP/ 00408/2020. AA acknowledges support from the grant RYC2021‐033226‐I funded by MCIN/AEI/10.13039/501100011033 and the European Union NextGenerationEU/PRTR.

%% file: files/6_appendix_gillespie.tex
\section{Gillespie algorithm\label{sec:gillespie}}

The action of the algorithm in a general model can be described as follows. Given a Markovian model, there is only a finite set of events that can happen. The algorithm first extracts a random waiting time before the next event; then, it randomly chooses the event that happens at the end of that time. In our model, the possible events are the following:
\begin{itemize}
    \item A node is infected by the black worm from a neighbour.
    \item A node is infected by the white worm from a neighbour.
    \item An user updates the device, removing the worm(s).
    \item A white worm becomes active.
    \item A white worm autonomously updates the device.
\end{itemize}

The first two events define an induced transition: a node can be infected only if it has a neighbour which is already infected. At a given state of the system, the rate at which an infection event happen is given by the product of the infection rate $\beta$ and the number of links between an infected node and an uninfected node. 

The other events are spontaneous transitions from one compartment to another, and their total rate in a given state will depend on the number of nodes in the initial compartment. For example, the rate at which a white worm becomes infectious will be given by $\epsilon (N_{D}+N_{D_B})$, where $N_D$ is the number of devices with only a dormant white worm and $N_{D_B}$ is the number of devices with also a black worm infection.

When the simulation begins, the algorithm first computes the total rate of the events, as the sum of all the rates of the possible events. It then extracts the waiting time for the next event from an exponential distribution with a rate equal to the total rate. The event which occurs is chosen randomly with probability proportional to the rate of the corresponding event. After the events, the rates are updated due to the new configuration, and the process is repeated. The process ends when the simulation time is more than a chosen $t_{max}$, or there are no more events that can happen. In the case of our model, the simulation will always stop since the devices eventually get either:
\begin{enumerate}
    \item All protected.
    \item Composed of a mixed population of protected, infected by the black worm only, and vulnerable but only connected to protected nodes.
\end{enumerate}
From both of these conditions, no other event can happen.

%% file: files/7_appendix_mean_field.tex
\section{Mean-field equations\label{app:meanfield}}
The equations of the model that define the mean-field approximation are 
\begin{equation}
\begin{cases}
&\dot{\rho}_k^V = - \beta_B \rho_k^V k \Theta_B - \beta_W \rho_k^V k \Theta_W, \\
&\dot{\rho}_k^B = \beta_B \rho_k^V k \Theta_B - \beta_W \rho_k^B k \Theta_W, \\
&\dot{\rho}_k^D = \beta_W \rho_k^V k \Theta_W - \beta_B \rho_k^D k \Theta_B - \epsilon \rho_k^D - \gamma_p \rho_k^D,  \\
&\dot{\rho}_k^{D_B} = \beta_B \rho_k^D k \Theta_B + \beta_W \rho_k^B k \Theta_W - \epsilon \rho_k^{D_B}  - \gamma_p \rho_k^{D_B},  \\
&\dot{\rho}_k^W = \epsilon \rho_k^D - \beta_B \rho_k^W k \Theta_B - \mu \rho_k^W, \\
&\dot{\rho}_k^{W_B} = \epsilon \rho_k^{D_B} + \beta_B \rho_k^W k \Theta_B - \mu \rho_k^{W_B},  \\
&\dot{\rho}_k^P = \mu \rho_k^{W_B} + \mu \rho_k^W + \gamma_p \rho_k^{D_B} + \gamma_p \rho_k^D,
\end{cases}
\end{equation}
where 
\begin{equation}
\begin{cases}
\Theta_B &= \sum_{k'} \frac{k' P(k')}{\langle k \rangle} (\rho_{k'}^B +\rho_{k'}^{D_B} + \rho_{k'}^{W_B}),  \\
\Theta_W &= \sum_{k'} \frac{k' P(k')}{\langle k \rangle} (\rho_{k'}^W + \rho_{k'}^{W_B}).
\end{cases}
\end{equation}